\documentclass[10 pt,aps,prl,twocolumn,superscriptaddress,showpacs,longbibliography]{revtex4-1}

\usepackage[utf8]{inputenc}
\usepackage{graphics}
\usepackage{epsfig}
\usepackage{dcolumn}
\usepackage{bm}
\usepackage{amsmath}
\usepackage{amsfonts}
\usepackage{latexsym}
\usepackage{amssymb}
\usepackage[normalem]{ulem}
\usepackage{color}
\usepackage{hyperref}

\newcommand{\g}[1]{{\bf #1}}
\newcommand{\ccs}{CeCu$_2$Si$_2$}
\newcommand{\ce}{CeCoIn$_5$}

\begin{document}

\title{Correlation-driven multigap $d$-wave superconductivity in Anderson lattice model}

\author{Marcin M. Wysoki\'nski}
\email{marcin.wysokinski@uj.edu.pl}

\affiliation{Marian Smoluchowski Institute of Physics$,$ Jagiellonian University$,$ 
ulica prof. S. Łojasiewicza 11$,$ PL-30-348 Krak\'ow$,$ Poland}

\author{Jan Kaczmarczyk}
\email{jan.kaczmarczyk@ist.ac.at}

\affiliation{Marian Smoluchowski Institute of Physics$,$ Jagiellonian University$,$ 
ulica prof. S. Łojasiewicza 11$,$ PL-30-348 Krak\'ow$,$ Poland}

\affiliation{Institute of Science and Technology Austria$,$ Am Campus 1$,$ A-3400 Klosterneuburg$,$ Austria}

\author{Jozef Spa\l ek }
\email{ufspalek@if.uj.edu.pl}

\affiliation{Marian Smoluchowski Institute of Physics$,$ Jagiellonian University$,$ 
ulica prof. S. Łojasiewicza 11$,$ PL-30-348 Krak\'ow$,$ Poland}

\date{\today}

\begin{abstract}
We present the full Gutzwiller-wave-function solution of the Anderson lattice model in two dimensions that leads 
to the correlation-driven multigap superconducting (SC)
ground state with the dominant $d_{x^2-y^2}$-wave symmetry. The results are consistent
with the principal properties of the heavy-fermion superconductor \ce. We regard the pairing mechanism as
universal and thus applicable to 
other Ce-based heavy-fermion compounds. Additionally, a gain in kinetic energy in the SC state takes place, as is
also the case for high-temperature superconductors.
\end{abstract}

\pacs{74.70.Tx, 74.20.-z, 71.27.+a, 74.20.Mn}
%

\maketitle{\it Introduction.} Starting from the first observation 
of superconductivity (SC) in the heavy fermion system (HFS) \ccs 
\cite{Steglich1979,Steglich2005}, there is an ongoing discussion 
concerning the microscopic mechanism(s) of pairing in this class 
of {\it unconventional superconductors}~\cite{Riseb2004,
Scalapino2012,dyke2014,Thompson2012}. At present, 
over thirty of HFS are known to be superconducting~\cite{Pfleiderer2009} 
and majority of them exhibit universal features 
of the electronic structure, that has their source in the strong interelectronic
correlations. Thus the assumption, that also the pairing may have
a common, non-phononic microscopic origin 
is widely accepted~\cite{Mathur1998,Riseb2004,Miyake2007,Scalapino2012,
dyke2014,Monthoux2007,Si2015,Tremblay2015}. Associated with this is the fundamental
question whether the strong correlations can also be the source of
pairing in those systems, in the same fashion, as they are for the appearance
of heavy quasiparticle states and non-trivial magnetic properties~\cite{Doradzinski1997,Doradzinski1998}.
Such a single approach might be regarded  as providing a class of universality defined by the type
of  many-particle theoretical  model. The purpose of this work is to present an affirmative 
answer to such a program by solving Anderson lattice model (ALM) with the full
Gutzwiller wave function for the SC state. The quasiparticle properties
in the normal state have been discussed elsewhere~\cite{Wysokinski2015}. 

We regard the two-orbital ALM as providing the most important 
electronic features of HF Ce-based compounds with 
an orbitally nondegenerate $f$ states (see e.g. Refs. 
\cite{Tremblay2015,Doradzinski1997,Doradzinski1998,Wysokinski2015}). 
Previous considerations of pairing in HFS of a purely 
electronic origin have involved related 
models such as the Kondo lattice model~\cite{Flint2010,
Pruschke2013,Fabrizio2013,Fabrizio2014}, the Anderson-Kondo 
lattice model~\cite{Spalek1988,Weber2008,Howczak2013,
Masuda2013} or ALM with the Falicov-Kimball term, the presence 
of which enhances the valence-fluctuation-induced pairing~\cite{Miyake2007}. 
Also, there exists a variational approach to the 
intrinsic SC pairing in ALM~\cite{Masuda2015}, with the 
s-wave character of SC gap.
A separate class of models has been based on the spin-fluctuation 
pairing among uncorrelated or weakly correlated electrons~\cite{Miyake1986,Monthoux2007, 
Scalapino2012}. 

Here we address the  mechanism of the 
heavy fermion superconductivity in ALM  driven by the
strong electronic correlations among $f$-orbital states and induced by 
the local repulsive $f$-$f$ Coulomb interaction, 
which represents by far the largest energy scale in the system. 
By incorporating this interaction into the full Gutzwiller wave function (GWF) solution
we obtain a stable SC state with two nodal gaps with the dominating 
$f$-$f$ and a weaker $f$-$c$ (conduction) electron real-space pairings, the former of
prevalent $d_{x^2-y^2}$ symmetry.
In this manner, the appearance of the superconducting pairing is proved as solely 
due to the electron correlations within one of the most representative methods of treating the correlated model
systems.
Such SC state reflects the principal features
exhibited by the canonical, quasi-two-dimensional 
\cite{Hall2001,Settai2001} HFS  \ce, for which the signatures of 
the nodal~\cite{Movshovich2001,Kohori2001,Curro2001,Aoki2004,Park2005,Stock2008} 
$d_{x^2-y^2}$~\cite{Izawa2001,Eskildsen2003,Weickert2006,
Vorontsov2006,Eremin2008,An2010,Rourke2005,allan2013,Zhou2013,dyke2014} 
SC of multigap character~\cite{Rourke2005,allan2013,Zhou2013,dyke2014} 
were reported. Moreover, as in our approach the leading SC component emerges from 
$f$ electrons, which are also responsible for the 
magnetic properties in HFS, our model accounts in a natural manner for 
the common origin of magnetic and SC orderings~\cite{Kenzelmann2008}.

{\it Model and method.}
Our starting point is the Anderson lattice model 
\begin{equation}
\begin{gathered}
 \mathcal{\hat H}={\sum_{\g i,\g j,\sigma}} 
 t_{\g i\g j}\hat c_{\g i\sigma}^\dagger\hat c_{\g j\sigma}
-\mu\sum_{\g i,\sigma}\hat n^c_{\g i\sigma}+ 
(\epsilon_f-\mu)\sum_{\g i,\sigma}\hat n_{\g i\sigma}^f\\+
U\sum_{\g i} \hat n_{\g i\uparrow}^f \hat n_{\g i\downarrow}^f+
\sum_{\g i,\g j,\sigma}(V_{\g i\g j}\hat f_{\g i\sigma}^\dagger
\hat c_{\g j\sigma}+\rm{H.c.}),
 \label{ALM}
\end{gathered}\vspace{-0.2cm}
\end{equation}
on a two-dimensional (2D), translationally invariant square lattice, 
with the chemical potential $\mu$, and with the usual notation~\cite{Wysokinski2015}. 

Ground-state properties of (\ref{ALM}) have been obtained variationally by means of 
a novel {\it Diagrammatic Expansion technique for the Gutzwiller Wave Function} (DE-GWF) 
~\cite{Wysokinski2015,buenemann2012,Kaczmarczyk2013,Kaczmarczyk2014,
Kaczmarczyk2014P,Kaczmarczyk2015}. The main advance introduced by this 
method is an accurate  evaluation of the expectation values
with the full Gutzwiller wave function (GWF).
The standard Gutzwiller Approximation (GA)~\cite{Rice1985}, and its subsequent statistically 
consistent elaboration (SGA)~\cite{Wysokinski2014,Wysokinski2014R,Wysokinski2015R}, are  reproduced 
in the zeroth order of our expansion and shown to 
include only the local correlations which do not lead to any SC solution. 
DE-GWF method accounts for nonlocal correlations being
the essential factor leading to the appearance of the SC state.
This particular feature of our approach is shared with the 
variational Monte Carlo (VMC) approach \cite{Edegger2007,Watanabe2015}. However, our technique 
does not suffer from the system finite-size limitation and is computationally efficient. 
In general, DE-GWF reproduces the VMC results with an improved accuracy, 
as has been shown previously on the example of SC solutions for both the Hubbard~\cite{Kaczmarczyk2013} 
and the $t$-$J$~\cite{Kaczmarczyk2014} models.

The Gutzwiller wave function, $|\psi_G\rangle$, is constructed from its uncorrelated correspondant, 
$|\psi_0\rangle$, by projecting out fraction of the local 
double $f$-occupancies by means of the Gutzwiller operator, i.e.,
$| \psi_G \rangle\equiv\mathcal{\hat P}_G|\psi_0 \rangle\equiv \prod_{\g i}
\mathcal{\hat P}_{G;\g i}| \psi_0 \rangle$.
The operator $\mathcal{\hat P}_{G;\g i}$ is defined by  
$\mathcal{\hat P}_{G;\g i}^\dagger\mathcal{\hat P}_{G;\g i}\equiv{\bf1} +x\hat d_{\g i}^{HF}$~\cite{Gebhard1990}, 
where $x$ is a variational parameter and $\hat d_{\g i}^{HF}\equiv\hat 
n^{HF}_{\g i\uparrow}\hat n^{HF}_{\g i\downarrow}=(\hat n_{\g i\uparrow}^f 
-n_{0f})(\hat n_{\g i\downarrow}^f -n_{0f})$ is relative to the Hartree-Fock (HF) double 
occupancy operator, with $n_{0f}\equiv\langle\hat n^f_{\g i\sigma}\rangle_0$. 
Hereafter we use a shortened notation ${\langle...\rangle_{G(0)}\equiv\frac{\langle
\psi_{G(0)}|...|\psi_{G(0)} \rangle}{\langle\psi_{G(0)}|\psi_{G(0)} \rangle}}$.

Formally, the expectation value with GWF for any product operator, 
$\hat{\mathcal{O}}_{\g i(\g j)}$, acting on site $\g i$ 
(and $\g j$), can be expanded in a power series in $x$
{\small
\begin{equation}
\begin{gathered}
 \langle \hat{\mathcal{O}}_{\g i (\g j)} \rangle_G\!=\!
 \Big\langle\hat{{\mathcal{O}}}^G_{\g i(\g j)}\prod_{\g l\neq\g i (\g j)} 
 \mathcal{\hat P}_{G;\g l}^2 \Big\rangle_0\!\!
 =\!\sum_{k=0}^\infty\frac{x^k}{k!}\!{\sum_{\g l_1,...,\g l_k}\!\!\!}'\!
 \langle\hat{{\mathcal{O}}}^G_{\g i(\g j)}\hat d^{HF}_{\g l_1,...,\g l_k}\rangle_0,
 \label{expe}
\end{gathered}
 \end{equation}}
where we have defined ${\hat{\mathcal{O}}^G_{\g i(\g j)}}\equiv 
\mathcal{\hat P}^\dagger_{G;\g i}(\mathcal{\hat P}^\dagger_{G;\g j})\hat{\mathcal{O}}_{\g i(\g j)}
(\mathcal{\hat P}_{G;\g j})\mathcal{\hat P}_{G;\g i}$ and 
$\hat d^{HF}_{\g l_1,...,\g l_k}\equiv\hat d^{HF}_{\g l_1}\cdots\hat d^{HF}_{\g l_k}$. 
The primed summation denotes the following restrictions: $\g l_p\neq\g l_{p'}$, 
and $\g l_p\neq\g i,\g j$ for all $p$ and $p'$. 
Power series in $x$ allows for a systematic incorporation of the long-range 
correlations among $k$ non-local sites ($\g l_1,...,\g l_k$) and the 
local ones ($\g i,\g j$). For $k=0$ we reproduce the results of GA, where only local
sites are projected. 
In this manner, the number of non-local correlated sites 
taken into account is the expansion parameter.
The resulting expectation values with $|\psi_0\rangle$ 
can be evaluated by applying the Wick's theorem. 
The products of the two-operator contractions can be visualized as 
diagrams, with sites and two-operator averages playing the role of 
vertices and lines, respectively~\cite{Kaczmarczyk2014,Wysokinski2015}. 
Here we allow for both the paramagnetic (PM) and SC contractions 
(lines) defined respectively as
{\small
\begin{equation}
\begin{gathered}
 P_{\g{l},\g{l'}}^{\alpha,\beta}\equiv \langle \hat \alpha_{\g{l}\sigma}^\dagger
 \hat \beta_{\g{l'}\sigma} \rangle_0-\delta_{\alpha f}\delta_{\beta f}
 \delta_{\g l{\g l'}}n_{0f}, \  \ \   
 S_{\g{l},\g{l'}}^{\alpha,\beta}\equiv \langle \hat \alpha_{\g{l}\sigma}^\dagger
 \hat \beta^\dagger_{\g{l'}\bar\sigma} \rangle_0,
 \label{lines}
 \end{gathered}
 \end{equation}}
where $\alpha,\beta\in \{c,f\}$ and $\g l,\g l'$ are the lattice indices. 
We have accounted for SC contractions, $S_{\g{l},\g{l'}}^{\alpha,\beta}$, 
which lead to the gap of $d$-wave symmetry, with the nodal
 lines along the diagonal directions.
 For an infinite lattice, we introduce a real space cutoff. Namely, we consider 
 only the lines limited by the distance $|\g l -\g l '|^2\equiv (l_x-l'_x)^2+
 (l_y-l'_y)^2\leq10$ (measured in lattice constants).  
The resulting expectation value of the Hamiltonian can be expressed by the 
diagrammatic sums (for details see Ref.~\cite{Wysokinski2015}), and in this  
manner, it depends explicitly on the variational parameter $x$, 
the correlation functions (lines) (\ref{lines}) and on $n_{0f}$.

The iterative procedure for obtaining the physical ground state of (\ref{ALM}) is as follows: \\
1. $\langle\mathcal{\hat H}\rangle_G$ is evaluated diagrammatically for selected $|\psi_0\rangle$.\\
2. $\langle\mathcal{\hat H}\rangle_G$ is minimized with respect to $x$. \\
3. The effective single particle Hamiltonian $\mathcal{\hat H}^{\rm eff}$ for the uncorrelated 
wave function $|\psi_0\rangle$ is determined.\\ 
4. New trial $|\psi_0'\rangle$ is obtained as the ground state of $\mathcal{\hat H}^{\rm eff}$. 
The points 1-4 are repeated in a self-consistent loop until a satisfactory convergence, i.e., 
the condition $|\psi_0\rangle=|\psi_0'\rangle$ is reached to a desired accuracy, typically $10^{-6}$.
The details of the method for the normal state of ALM are thoroughly discussed in Ref.~\cite{Wysokinski2015}.

The effective single-particle Hamiltonian $\mathcal{\hat H}^{\rm eff}$ for the uncorrelated 
wave function $|\psi_0\rangle$ is determined from the condition that its optimal expectation value 
coincides with $\langle\mathcal{\hat H}\rangle_G$. This leads to the condition 
{\small
\begin{equation}
\begin{gathered}
\delta\langle \mathcal{\hat H}^{\rm eff}\rangle_0 (P_{\g{l},\g{l'}}^{\alpha,\beta},
S_{\g{l},\g{l'}}^{\alpha,\beta},n_{0f})= 
\delta\langle \mathcal{\hat H}\rangle_G (P_{\g{l},\g{l'}}^{\alpha,\beta},
S_{\g{l},\g{l'}}^{\alpha,\beta},n_{0f})\\
=\sum_{\g l,\g{l'}} \Big{(}\frac{\partial \langle \mathcal{\hat H}\rangle_G}
{\partial P_{\g{l},\g{l'}}^{\alpha,\beta}}\delta P_{\g{l},\g{l'}}^{\alpha,\beta} 
 +\frac{\partial \langle \mathcal{\hat H}\rangle_G}{\partial S_{\g{l},\g{l'}}^{\alpha,\beta}}
 \delta S_{\g{l},\g{l'}}^{\alpha,\beta}\Big{)}
 +\frac{\partial \langle \mathcal{\hat H}\rangle_G}{\partial n_{0f}}\delta n_{0f}.
\end{gathered}
\end{equation}}
Explicitly, the effective single-particle Hamiltonian reads,
{\small
\begin{equation}
\begin{gathered}
 \mathcal{\hat H}^{\rm eff}=\sum_{\g i,\g j,\sigma}\Big [
 t_{\g i\g j}^{cc}\hat c_{\g i\sigma}^\dagger\hat c_{\g j\sigma}
+t_{\g i \g j}^{ff}\hat f_{\g i\sigma}^\dagger\hat f_{\g j\sigma}
+t_{\g i\g j}^{cf}(\hat c_{\g i\sigma}^\dagger\hat f_{\g j\sigma}+{\rm H.c.})\Big]\\
+\sum_{\g i,\g j,\sigma}\Delta_{\g i\g j}^{cf}
(\hat c_{\g i\sigma}^\dagger\hat f_{\g j\bar \sigma}^\dagger 
+{\rm h.c.})+\sum_{\g i,\g j}\Delta_{\g i\g j}^{ff}
(\hat f_{\g i\uparrow}^\dagger\hat f_{\g j\downarrow}^\dagger +{\rm H.c.}),
\label{heff}
\end{gathered}
\end{equation}}
with the effective microscopic parameters determined by 
\begin{equation}
 t_{\g i\g j}^{\alpha\beta}= \frac{\partial \langle\mathcal{H}\rangle_G}
  {\partial P_{\g{i},\g{j}}^{\alpha,\beta}}, \ \
  \Delta_{\g i\g j}^{\alpha\beta}= \frac{\partial \langle\mathcal{H}\rangle_G}
  {\partial S_{\g{i},\g{j}}^{\alpha,\beta}}, \ \ t_{\g i\g i}^{ff}= 
  \frac{\partial \langle\mathcal{H}\rangle_G}
  {\partial n_{0f}}.\label{teff}
\end{equation}
Parenthetically, as GWF introduces correlations within the 
$f$ orbital only, there is no effective pairing 
between $c$-electrons, since there are no $S_{\g{i},\g{j}}^{c,c}$ lines in the diagrams 
visualizing the Wick's contractions in (\ref{expe}).

In the momentum space, the effective 
Hamiltonian can be reformulated in the Bogoliubov - de Gennes - Nambu form 
\cite{degennes1966}
\begin{equation}
\begin{gathered}
 \mathcal{\hat H}^{\rm eff}=\sum_{\g{k}}  \Psi_\g{k}^\dagger
\begin{pmatrix}
 \epsilon_{\g{k}}^{cc} &0&\epsilon_{\g{k}}^{fc}&\Delta_{\g{k}}^{fc} \\
0& -\epsilon_{\g{k}}^{cc}&\Delta_{\g{k}}^{fc}&-\epsilon_{\g{k}}^{fc}\\
\epsilon_{\g{k}}^{fc}&\Delta_{\g{k}}^{fc} &\epsilon_{\g{k}}^{ff}&\Delta_{\g{k}}^{ff} \\
\Delta_{\g{k}}^{fc}&-\epsilon_{\g{k}}^{fc}&\Delta_{\g{k}}^{ff}& -\epsilon_{\g{k}}^{ff} \\ 
\end{pmatrix}
\Psi_\g{k},
\label{hk}
\end{gathered}
\end{equation}
where we have defined $\Psi^\dagger\equiv(\hat c_{\g{k}\uparrow}^\dagger,
\hat c_{\g{-k}\downarrow},\hat f_{\g{k}\uparrow}^\dagger,\hat f_{\g{-k}\downarrow})$
and ${\epsilon_{\g{k}}^{\alpha\beta}(\Delta_{\g{k}}^{\alpha\beta})=(1/L)\sum_{\g i\g j}
t_{\g i\g j}^{\alpha\beta}(\Delta_{\g i\g j}^{\alpha\beta})e^{i(\g{i}-\g{j})\g k}}$, where
$L$ is the number of lattice sites. 
Hamiltonian (\ref{hk}) can be easily  diagonalized by the Bogoliubov-type of transformation and thus  
the new lines and $n_{0f}$ 
determining the ground state of $ \mathcal{\hat H}^{\rm eff}$ can be obtained, i.e.,
\begin{equation}
\begin{split}
 P_{\g{i},\g{j}}^{\alpha,\beta}\Big(S_{\g{i},\g{j}}^{\alpha,\beta}\Big)&=\frac{1}{L}\sum_{\g k}
 \langle\hat\alpha_{\g k}^\dagger\hat\beta_{\g k}\rangle_0
 \Big(\langle\hat\alpha_{\g k}^\dagger\hat\beta^\dagger_{\g k}\rangle_0\Big)e^{i(\g{i}-\g{j})\g k},\\
 n_{0f}&=\frac{1}{L}\sum_{\g k}
 \langle\hat\alpha_{\g k}^\dagger\hat\beta_{\g k}\rangle_0.
 \label{newlines}
 \end{split}
 \end{equation}
 
To determine the equilibrium properties of ALM, the system of Eqs. 
(\ref{teff}) and (\ref{newlines}) is solved self-consistently, 
together with the minimization of $\langle \mathcal{\hat H} \rangle_G$ 
with respect to $x$~\cite{buenemann2012,Kaczmarczyk2013,Kaczmarczyk2014,
Kaczmarczyk2014P,Kaczmarczyk2015}. 
We also adjust the chemical potential as the total filling $n\equiv2
\langle\hat n_{\g i\sigma}^f+\hat n_{\g i\sigma}^c\rangle_G$ is fixed. 
Finally, the physical ground state energy of the system is obtained as
$E_G=\langle \mathcal{\hat H} \rangle_G|_0/L+n|_0\mu$, where $|_0$ denotes 
the equilibrium value. 
We define also the $f$-orbital filling as $n_f=2\langle\hat n_{\g i\sigma}^f\rangle_G$.

Our variational scheme enables us to determine both $\mathcal{\hat H}^{\rm eff}$ 
and $|\psi_0\rangle$  and in turn, the renormalized SC 
order parameters
\begin{equation}
 \Delta^{\alpha\beta}_{G;i_x-j_x,i_y-j_y}\equiv\langle \hat\alpha_{\g i}^\dagger
 \hat\beta_{\g j}^\dagger\rangle_G=\langle \mathcal{\hat P}_G\hat\alpha_{\g i}^\dagger
 \hat\beta_{\g j}^\dagger\mathcal{\hat P}_G\rangle_0.
 \label{delta}
\end{equation}
Note that although, there is no pairing term between the $c$ electrons in (\ref{heff}), 
the corresponding SC order parameters are nonzero.

\begin{center}\vspace{-0.1cm}
  \begin{figure}[t]
   \includegraphics[width=0.5\textwidth]{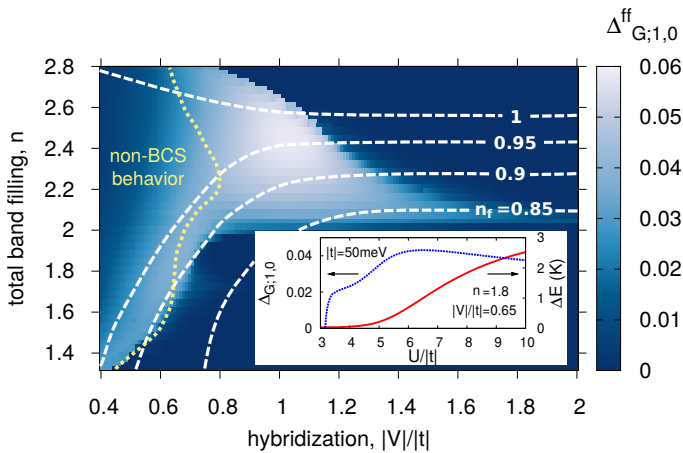}   
   \caption{Phase diagram on total band filling -- hybridization 
   magnitude plane, with the dominant, $d_{x^2-y^2}$-wave 
   superconducting order parameter for $f$ electrons, 
   $\Delta_{G;1,0}^{ff}$ (see main text). 
   The dashed lines mark the selected $f$-orbital isovalents.
   The dotted line singles out the {\it non-BCS} region defined 
   by the gain in the kinetic energy, $\Delta E_{kin}>0$ in SC state.
   For the selected point ($n=1.8$, $|V|/|t|=0.65$) we show in the inset 
   that SC appears with the increasing $f$-$f$ interaction.}
  \label{fig1}
  \end{figure} 
 \end{center}\vspace{-0.7cm}

{\it Results.}
We have selected the starting microscopic parameters realistic 
for the Ce-based HFS, namely $c$-electrons with  
nearest-neighbor hopping $t$ and the second 
nearest-neighbor hopping, $t'=0.25|t|$, $f$-electron atomic level 
position  $\epsilon_f=-3|t|$, and $f$--$f$ Coulomb repulsion, $U=10|t|$. 
The bare $c$--$f$ hybridization with the amplitude $V$ is considered 
as a variable and of the nearest-neighbor origin.
Physical energies are obtained by assuming $|t|=50$ meV 
(see e.g.~\cite{allan2013, dyke2014}). 
We have restricted our analysis to the $f$ orbital filling not exceeding 
unity or slightly larger, $n_f\lesssim1.05$, to include possible local 
$f^{3+}$, $f^{4+}$, and $f^{2+}$ configurations. 
The diagrammatic expansion, if not stated otherwise, has been carried out 
up to the third order, $k=3$. 

In Fig.~\ref{fig1} we present the phase diagram characterized by the leading value of 
the $f$-$f$ electron $d_{x^2-y^2}$ component of the order parameter, 
$\Delta^{ff}_{G;1,0}$ on the total filling -- hybridization, $n$--$|V|$ plane, with 
the marked isovalents (dashed lines) by the $f$ orbital occupation number, $n_f$.
It can be inferred that correlations for $f$-orbital filling in the range 
$n_f\lesssim 0.8-0.85$ are too weak to lead to robust SC state. 
In the inset to Fig.~\ref{fig1} (for the selected point on the diagram) 
we show explicitly the appearance of $\Delta^{ff}_{G;1,0}$ and the associated 
condensation energy $\Delta E=E_{PM}-E_{SC}$ emerging with the increasing  
interaction amplitude, $U$. Here, we have defined $E_{PM}$ and $E_{SC}$ as the ground state 
energies for the normal and SC states, respectively. 

 \begin{center}\vspace{-0.18cm}
  \begin{figure}[t]
   \vspace{0.21cm}
   \includegraphics[width=0.46 \textwidth]{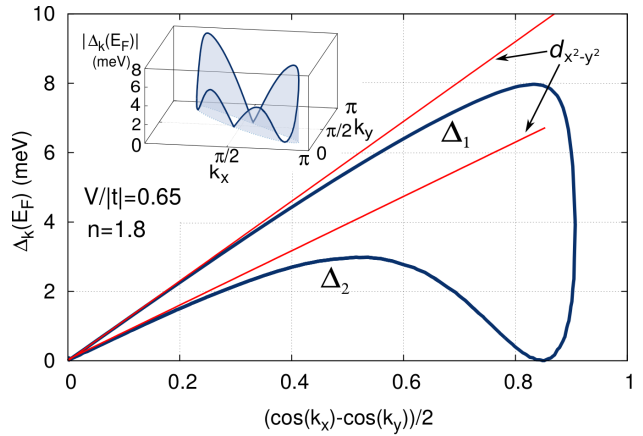}   
   \caption{ Angle dependence of the SC gaps at the Fermi 
   surface for selected parameters $n=1.8$ and 
   $|V|/|t|=0.65$. The larger gap, $\Delta_1$ follows $d_{x^2-y^2}$ dependence. 
   The inset shows the quarter of the 
   Brillouin zone with explicitly drawn absolute values of the SC gaps.
   }\label{fig5}
   \end{figure} 
 \end{center}\vspace{-0.8cm}

The order parameter $\Delta^{ff}_{G;1,0}$ is representative as it dominates practically always. 
In particular, in the singled out region, marked as 
{\it non-BCS}, the absolute values of the remaining order parameters 
$\Delta^{\alpha\beta}_{G;i_x-j_x,i_y-j_y}$ are less than 35\% of $\Delta^{ff}_{G;1,0}$.
In the {\it non-BCS} region, in distinction to that of BCS,  a gain 
in the kinetic energy appears with respect to PM state,
${\Delta E_{kin}\equiv E^{kin}_{PM}-E^{kin}_{SC} > 0}$. 
The kinetic energy comprises all the contributions to the ground state energy 
except the potential part, i.e., $E_{kin}\equiv E_G-\frac{U}{L}\langle \sum_{\g i} 
\hat n^f_{\g i\uparrow}\hat n^f_{\g i\downarrow} \rangle_G$.
In the {\it non-BCS} regime the density of states at the Fermi level (not shown explicitly) in PM state  
is significantly enhanced signaling heavy quasiparticle masses, and suggesting that this parameter 
range is appropriate for the HFS description. 
On this basis, we predict that the gain in the kinetic energy in the SC state 
for the Ce-based heavy fermion superconductors, and for specifically addressed 
here \ce, is the next feature 
shared with the high-temperature superconductors, in addition to the 
$d$-wave symmetry of the order parameter,  
the competition of SC with antiferromagnetism, and emergence of pseudogap~\cite{Zhou2013}.

\ce\ represents a thoroughly studied SC system
\cite{Pfleiderer2009,Thompson2012} among  
Ce-based superconductors. Unprecedentedly 
among HFS, for this compound the gap function was directly observed  
and characterized~\cite{Zhou2013,allan2013,dyke2014}. 
To compare our results to those of \ce\ 
we have selected the representative point on the phase 
diagram (cf. Fig.~\ref{fig1}), 
determined by the values $n=1.8$ and $|V|\simeq 0.65$.
The total filling $n$ has been chosen so that
the Fermi level is placed  in the energy below  
the middle point of the hybridization gap 
\cite{allan2013,Zhou2013}. 
In turn, hybridization amplitude $V$ has been adjusted to obtain the  
value of the condensation energy ($\Delta E\simeq 2.3 K$ - cf. Fig.~\ref{fig1}, inset) 
corresponding to the SC critical temperature of \ce.  

We obtain a nodal pairing with the 
two distinct gaps at the Fermi surface, in accord with the findings for \ce\ \cite{Zhou2013,allan2013,dyke2014,Rourke2005}. 
In the Fig.~\ref{fig5}, we have shown explicitly their angular dependence (and shape - cf. inset) 
in the quarter of the first Brillouin zone.
The larger gap, $\Delta_1$ follows pure $d_{x^2-y^2}$-symmetry dependence as
suggested for \ce\ by various experiments cf. e.g. Refs.~\cite{Zhou2013,allan2013,dyke2014}.
 \begin{center}\vspace{-0.1cm}
  \begin{figure}[t]\vspace{0.24cm}
   \includegraphics[width=0.435\textwidth]{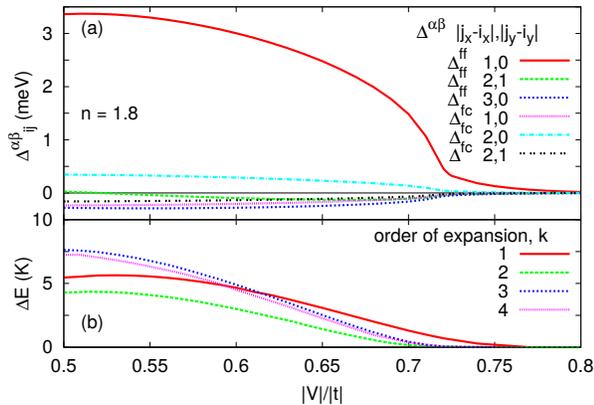}   
   \caption{(a) Effective gap components with the dominant
   $d_{x^2-y^2}$-wave $f$-$f$ part. Components with $\Delta^{\alpha\beta}_{\g i \g j}/
   \Delta^{ff}_{1,0}\!<\!5\%$
   are not included. (b) Condensation energy for SC state in consecutive 
   orders of the expansion.  In all plots $n=1.8$.}\label{fig4}
   \end{figure} 
 \end{center}\vspace{-0.8cm}

  \begin{center}\vspace{-0.1cm}
  \begin{figure}[t]   
   \includegraphics[width=0.45\textwidth]{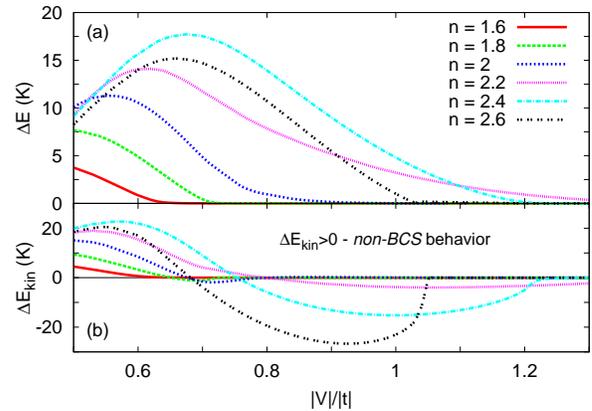}
   \caption{(a) The value of the condensation energy 
   $\Delta E$ for the fixed values of the total filling $n$. 
   (b) Kinetic energy difference, $\Delta E_{kin}$ between normal and 
   SC states. Behavior, when $\Delta E_{kin}>0$ is opposite to that 
   predicted by the BCS theory.}\label{fig3}
   \end{figure} 
 \end{center}\vspace{-0.8cm}
 
To track the leading contribution to the pairing, in Fig.~\ref{fig4}a we display the effective SC pairing components 
(lines) along the selected direction of constant band filling, $n=1.8$. The dominant pairing amplitude, 
$\Delta^{ff}_{1,0}$ describes the  
$f$-$f$ pairing part and is of the $d_{x^2-y^2}$-wave symmetry. 
The remaining $f$-$f$ and $f$-$c$ pairing components constitute 
less than 15\% of  $\Delta^{ff}_{1,0}$. Such circumstance 
explains in a natural manner the common origin of magnetic and 
SC orderings~\cite{Kenzelmann2008} in \ce\ as the former is 
 associated mostly with $f$-electrons. 
Additionally, in Fig.~\ref{fig4}b we present the convergence of our results with the order
$k$ of the expansion on example of the condensation energy for 
$n=1.8$. Absence of any considerable difference between the $k=3$ and $k=4$ plots
proves that the convergence is already reached for $k=3$.
 
 In Fig.~\ref{fig3}a  we plot the condensation energy, $\Delta E$ for different $n$ values. 
The energy which should be of the order of the transition
temperature is reasonable, especially 
in the limit of the total filling, $n<2$. Nonetheless, $\Delta E$ in 
our model can be suppressed by introducing an onsite contribution to hybridization, here 
considered of a purely intersite form. 
In Fig.~\ref{fig3}b we plot the gain in kinetic energy 
in SC state. 
The {\it non-BCS} state appears for lower values of the hybridization amplitude $|V|$.

 {\it Summary.}
 The Anderson lattice model has been solved diagrammatically with a full Gutzwiller wave function.
This leads to the generic correlation-driven unconventional superconductivity. 
The SC state exhibits principal properties detected in a clear manner in \ce\ \cite{Zhou2013,allan2013,dyke2014}, 
as well as can be expected to appear also in other Ce-based superconductors~\cite{Pfleiderer2009,Scalapino2012}:
({\it i}) superconductivity is of multigap ($f$-$f$ and $f$-$c$ pairings) character; 
({\it ii}) the leading gap component is due to $f$-$f$ pairing and of $d_{x^2-y^2}$-wave symmetry; and 
({\it iii}) the condensation energy of a reasonable value, i.e. of the order of critical temperature.
Additionally, we also show that in a direct analogy to high-temperature superconductors,
heavy fermion systems can be characterized by the presence of non-BCS regime.

{\it Acknowledgements.} We are very grateful for stimulating 
discussions with J. B\"unemann. 
The work has been supported by the  
National Science Centre (NCN) under the Grant MAESTRO, 
No. DEC-2012/04/A/ST3/00342. 
JK acknowledges support from the People Programme (Marie Curie Actions) of the 
European Union's Seventh Framework Programme (FP7/2007-2013) 
under REA grant agreement n$^{\rm o}$ [291734]. 

 %

\end{document}